\documentclass[aps,pre,twocolumn,showpacs]{revtex4-1}
\usepackage{amsmath}
\usepackage{epsfig}
\usepackage{amssymb}
\usepackage{dcolumn}
\usepackage{graphicx}
\usepackage{dcolumn}
\usepackage{graphics}
\usepackage{color}

\begin{document}

\title{Dynamical generation of interwoven soliton trains by nonlinear emission in binary Bose-Einstein condensates}

\author{V.A. Brazhnyi$^1$, David Novoa$^{2,3}$,  Chandroth P. Jisha$^1$}
\affiliation{ $^1$Centro de F\'{\i}sica do Porto, Faculdade de Ci\^encias, Universidade do Porto, R. Campo Alegre 687, Porto 4169-007, Portugal;\\
$^2$Max Planck Institute for the Science of Light, G\"unther-Scharowsky str. 1, 91058 Erlangen, Germany;\\
$^3$Centro de L\'aseres Pulsados, CLPU. Edificio M3 - Parque Cient\'{i}fico. Calle del Adaja, s/n, Villamayor, ES-37185 Spain}

\date{\today}
\begin{abstract}
We propose a method for the generation of trains of alternating bright solitons in two-component Bose-Einstein condensates, using controlled emission of nonlinear matter-waves in the uncoupled regime with spatially-varying intra-species interaction and out-of-phase oscillations of the ground states in the trap.
Under this scheme, solitons are sequentially launched from the different components, and interact with each other through phase-independent cross-coupling. 
We obtain an analytical estimation of the critical condition for soliton emission using a geometric guiding model, in analogy with integrated optical systems.
In addition, we show how strong initial perturbations in the system can trigger the spontaneous generation of supersolitons, i.e. localized phonon-like excitations of the soliton trains. Finally, we demonstrate the controllable generation of slow and fast supersolitons by adding external localized potentials in the nonlinear region.

\end{abstract}


\maketitle

\section{Introduction}
The advent of Bose-Einstein condensates (BECs)~\cite{anderson95} paved 
the way for the generation of coherent atomic bursts with properties analog to those of optical laser beams. 
Such \emph{atom lasers}, which promise significant achievements in fields
like atomic interferometry~\cite{interferometer,michinel12}, have been proposed operating in both continuous~\cite{carpentier10} and pulsed~\cite{mewes97, carr04, rodas05} regimes, and with different outcoupling
mechanisms~\cite{guerin06, couvert08, gattobigio09}. Concretely,
atom lasers based on the emission of bright solitons~\cite{solitonBEC,perez98} have also been recently proposed~\cite{carr04, rodas05}. 
Their operation is essentially based on the nonlinear mechanism of modulational instability, 
which results in the generation of trains of atomic solitons~\cite{train02}. Thus, atomic soliton lasers display a fully reconfigurable directional emission owing to 
the presence of inhomogeneous (i.e., spatio~\cite{rodas05}-temporal~\cite{michinel12} dependent) interactions between the atoms in the ultracold cloud.

It is worth noticing that, to our best knowledge, the emission of localized matter-wave bursts have been 
mainly studied theoretically in the framework of one-component BECs~\cite{carr04,rodas05,carpentier10,michinel12}. However,
the recent demonstration of tunable interspecies interactions in Bose-Bose~\cite{thalhammer08} and Bose-Fermi mixtures~\cite{Best09} 
have opened new avenues for the generation and control of multicomponent or vector nonlinear mater-waves~\cite{vector_solitons}. In particular, new types of phonon-like 
excitations in binary atomic soliton chains, the so-called \emph{supersolitons} (SS), and a realization of the Newton's cradle using bright matter-wave solitons as building blocks have been theoretically proposed~\cite{novoa08}. One of the main drawbacks for the experimental observation of the latter phenomenology is the need for efficient generation of trains of interlaced solitons in both atomic components. While, to our best knowledge, such kind of nonlinear structures have not been accomplished in the laboratory yet, a theoretical proposal based on the temporal tailoring of Feshbach resonances has recently come out~\cite{feijoo2013}. 

The goal of the present work is twofold. First, we propose a feasible route for the dynamical generation and control of interwoven soliton trains in binary BECs. Our method relies on the \emph{one-by-one} emission of alternating matter-waves coming from binary BECs, in sharp contrast with the first experimental realization of atomic soliton trains in one-component BECs~\cite{train02}, where the soliton bursts were created at once via modulational instability induced by the magnetic tuning of the intra-atomic interactions from 
repulsive to attractive along the whole BEC cloud. In our case, the nonlinear outcoupling mechanism of the individual solitons is then based on the tailoring of spatially-inhomogeneous interactions between the ultracold atoms, which allows for the generation of unstable nonlinear surface states localized at the edge of the trapping potential. 
As we shall extensively discuss in this manuscript, such surface states can eventually destabilize giving rise to the release of atomic solitons out of the trap in a fully-controllable way.

Second, we demonstrate the performance of our scheme in a scenario that might be accomplished in the framework of current experiments, showing its potential to enable the first observation of elastic-like scattering between solitons belonging to different BEC components, as well as the generation of SS as a result of the collective transfer of linear momentum among all neighboring solitons in the interwoven trains.

In particular, in section II we introduce the physical model that describes the evolution of the binary BEC in the mean-field limit. In section III, we analyze the stationary states of the system, in order to get some insight into the nonlinear matter-wave emission process~\cite{rodas05}. As a matter of fact, we will demonstrate that the emission of matter-wave solitons is achieved when the particle density at the border of the trapping potential exceeds a certain critical threshold, which turns out not to be trivially linked to the critical number of particles that can be loaded in the trap in presence of attractive interatomic interactions~\cite{rodas05}. In a particular configuration including a super-Gaussian trap, we will discuss in section IV an analytical approach to the critical amplitude yielding to soliton emission, exploiting the formal analogies between this system and one-dimensional, step-index optical waveguides. Finally, in section V we will study the dynamical generation of 
interwoven soliton trains and subsequent excitation of SS on top of them. Thus, we will show how this issue can be fixed by appropriately tuning the interspecies interactions in space and providing a certain number of solitons in the train with an additional momentum by switching on an additional dipole potential~\cite{dipoletrap} in specific space-time locations. 

\section{Physical model}

The dynamics of \emph{cigar-shaped} two-component BECs, i.e., tightly confined in two transverse spatial dimensions ($y,z$) and weakly confined in the longitudinal direction ($x$) by a trapping potential  $V(x)$, can be accurately described in the mean-field limit by the following set of coupled Gross-Pitaevskii equations (GPE)~\cite{BECreview}
\begin{equation}
 \label{gpe}
 i\frac{\partial u_j}{\partial t} = - \frac{1}{2} \frac{\partial^2 u_j}{\partial x^2} + V(x)u_j-\sum_{k=1,2}a_{jk}|u_k|^2 u_j,
\end{equation}
where $u_j$ are the condensate wave-functions of the two atomic BECs ($j=1,2$), which can be constituted either by different atomic species or the same atomic species in different hyperfine states.  
The spatial variable $x$ and time $t$ are measured, respectively, in units of  $a_0=\sqrt{\hbar/m\omega_\bot}$ and $1/\omega_\bot$, being $\omega_\bot$ the angular frequency of the strong transverse confinement of the BEC cloud. Taking into regard the aforementioned normalizations, we define the number of particles as $N_j=\int |u_j|^2 dx$. 

To be concrete, the normalization factors ($\omega_\bot,a_0$) can be estimated for a possible experimental realization of the phenomenology to be discussed in this paper. Hence, we will consider parameters corresponding to condensates of $^{7}$Li atoms tightly confined in the harmonic trap with angular frequency $\omega_\bot\approx 5\cdot 10^3~rad/s$ \cite{train02}. Under these conditions, the transverse size of the ultracold cloud turns out to be $a_0\approx 1.3\ \mu m$. The normalized number of particles of each component, $N_j$, is related to the real number of atoms ${\cal N}_j$ through the relation $N_j=g {\cal N}_j$, where $g\propto a_{jj}/a_0$ displays values around $10^{-3}\div 10^{-4}$ in real experimental situations.

 On the other hand, the coefficients $a_{jk}$ are proportional to the scattering lengths, so that they characterize the strength of both intra-atomic ($j=k$) and inter-atomic ($j \neq k$) interactions. Moreover, depending on the sign, they model either attractive ($a_{jk} > 0$) or repulsive ($a_{jk} < 0$) interactions between the particles. In this work, we will also consider inhomogeneous interactions in space, modeled by spatially-varying $a_{jk}$ coefficients defined as follows
 
   \begin{equation}
     \label{scatt}
     a_{jk} = \left\{
	       \begin{array}{ll}
		 H(x_0)c_{jk}       &\mathrm{if\ } j=k; \\
		 H(x_{00})c_{jk}  &\mathrm{if\ } j \neq k,
	       \end{array}
	     \right.
   \end{equation}
where $H(x_0)$, $H(x_{00})$ are Heaviside functions and $c_{jk}$ are constants carrying the information about the strength of $a_{jk}$ (we have checked that the choice of smoother functions to model the spatial variation of the scattering lengths does not change qualitatively the results presented below). In general, the spatial coordinates $x_0$, $x_{00}$ will be different.
We must point out that the scattering lengths can be tuned in real experiments by means of either external magnetic~\cite{Inouye98} or optical fields~\cite{Fatemi00}.
In particular, using optical fields, spatial control of interatomic interactions down to the submicron resolution was recently demonstrated in a BEC of Yb atoms\cite{submicron}. Even though the experimental realization of the phenomenology presented here could be more involved owing to the need for a fine spatial control of the scattering length at different locations, the results of Ref. \cite{submicron} might pave the way for future improvements in this direction.
Without loss of generality, we consider the confining potential as a super-Gaussian function $V(x) = V_0 \exp\left[-(x/d)^{2l}\right]$, where $l$ is a positive integer and $2d$ defines the width of the potential. By taking $V_0<0$ and $l\gg1$ we can mimic a box-like potential well with depth $V_0$ and width $2d$, which, in combination with the sharp spatial variation of $a_{jk}$, turns out to be useful for analytical treatment as we will see below.

\section{Stationary problem}

We begin by solving the nonlinear eigenvalue problem associated with Eq.(\ref{gpe}). As we are interested in the 
analysis of the nonlinear matter-wave emission, let us consider, for the sake of simplicity, the scalar case where $c_{12}=c_{21}=0$ and $c_{jj}=c=1$. The existence of coupled vectorial stationary states in this system is beyond the scope of the present paper and will be reported elsewhere. Under the aforementioned conditions, we have two identical uncoupled equations for both condensate wave-functions, which can be cast in the following form
\begin{equation}
 \label{st_gpe_1}
 \mu U = - \frac{1}{2} \frac{\partial^2 U}{\partial x^2} + V(x)U - H(x_0) |U|^2 U.
\end{equation}
Here we search for stationary solutions of the form $u_{1,2}(x,t)=U(x)e^{-i\mu t}$, where $U(x)$ is the spatial profile of the localized solution and $\mu$ is the chemical potential. Similar systems were studied in detail previously~\cite{carpentier10,rodas05}. However, all those works assume the spatial region $x>x_0$ where nonlinear self-interactions are non-negligible, to be fixed with respect to the center of the trap.  In this work, we will show that changes in $x_0$ modify the structure of the eigenstates fundamentally, therefore affecting the dynamical behavior of the system itself. Furthermore, the choice of a super-Gaussian-type potential is essential for the control of the dynamics of generation of interwoven soliton trains because of its smooth boundary, as we will discuss in section IV. Hereafter, unless otherwise stated, we will fix the width of the potential ($2d$) and change the position of the nonlinear boundary considering always $x_0>d$.

\begin{figure}[h]
\epsfig{file=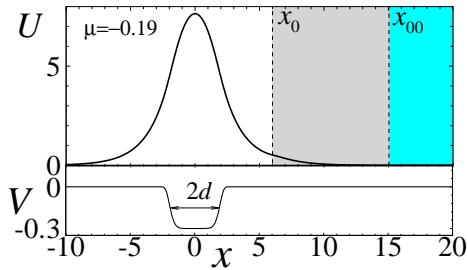,width=7cm}
\caption{ (Color online) Upper panel: Spatial profile $U(x)$ of the eigenstate featuring $N=N_{max}$ with $x_0=6$. Lower panel: Shape of the trapping potential with $V_0=-0.3$, $d=2$ and $l=4$. In the upper panel, shaded domains indicate the regions where just self-interactions (gray) or both self- and cross-interactions (blue) are active.}
\label{ini}
\end{figure}

\begin{figure}[h]
\epsfig{file=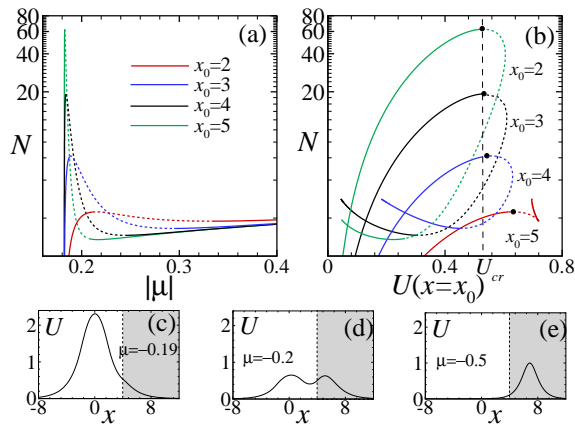,width=8cm}
\caption{(Color online)
Panel (a): Existence curves of the eigenmodes $N(|\mu|)$ in the scalar case ($c_{jk}=0$, $j\neq k$) for different positions of the nonlinear boundary: $x_0=2, 3,4$ and $5$. Solid and dashed lines correspond to the stable and unstable domains, respectively. Panel (b): Dependence of the amplitude of the solution at the boundary  $U(x=x_0)$ as a function of $N$. The  black points correspond to $N=N_{max}$.
Other parameters are: $c=1$, $V_0=-0.3$, $d=2$, $l=4$. Panels (c)-(e): Profiles of the stationary states for different chemical potentials $\mu$, in the case $x_0=4$. } \label{fig1}
\end{figure}

An example of stationary state is depicted in Fig.\ref{ini}, where the nonlinear interactions between particles are non-vanishing in the shaded regions. On the other hand, 
in Fig.\ref{fig1}(a), we display the number of particles $N$ of the corresponding \emph{nodeless} stationary states of the system, as function of the chemical potential $\mu$ (existence curves) for different positions of the nonlinear boundary $x_0$.

For $N\rightarrow 0$ (linear limit), the number of particles in the nonlinear region is small and so it does not affect the shape of the linear ground state of Eq.(\ref{st_gpe_1}). As a consequence, all the existence curves arise over a certain minimum  threshold of the chemical potential $|\mu_{min}|>0$, which is determined by both the width and the depth of the trap,  and corresponds to the eigenvalue of the ground state of $V(x)$ in the absence of nonlinear interactions (i.e., for $x_0\gg1$) (in the particular case $d=2$ shown in Fig.\ref{fig1} $\mu_{min}\approx-0.182$).

For larger $N$, however, the particle density at the nonlinear region becomes stronger and leads to the   
appearance of a local maximum $N=N_{max}$ in the existence curves [see Fig \ref{fig1}(a)]. Notice that the magnitude of $N_{max}$ grows as the position of the nonlinear boundary $x_0$ increases. Hereafter, for the case $N=N_{max}$ we will define the critical amplitude  $U_{cr}$ as the amplitude of the eigenstate  at the point $x=x_0$. As an illustrative example, in Fig.\ref{fig1}(c) we display the profile of the eigenstate at $N=N_{max}$ for $x_0=4$.
 
Further growth in $|\mu|$ transforms the \emph{stable} nonlinear solutions into \emph{unstable} surface modes localized at the boundary between the linear and nonlinear region $x_0$ [see Fig. \ref{fig1}(d)]. Similar behavior was predicted in symmetric layered optical systems~\cite{Akhmediev82}. All surface modes found in this system turn out to be unstable against small perturbations (represented by dashed lines in Fig. \ref{fig1}). This assertion has been checked by calculating the perturbational spectrum of the eigenvalue problem that results after linearization of Eq.(\ref{gpe}). The results of the linear stability analysis were further confirmed by direct numerical simulations of evolution. It is remarkable that such a behavior can be predicted by the Vakhitov-Kolokolov stability criterion~\cite{VK}, even though its validity is not rigorously justified in systems with inhomogeneous nonlinear responses. 

As it can be appreciated in Fig. \ref{fig1}(a), by further decreasing the chemical potential the surface mode decays into a stable soliton [see Fig. \ref{fig1}(e)]. Such solitary wave is localized almost entirely in the nonlinear region, yielding to the stable branch of the existence curve with growing $N(|\mu|)$, as $|\mu|\rightarrow \infty$. The point of this transformation corresponds to the local minimum of the existence curve $N=N_{min}$, which constitutes the minimum number of particles required to form a stable bright soliton. Noteworthy, for $x_0<2$ the unstable part of the existence curve shrinks and disappears, transforming $N(|\mu|)$ dependence into monotonously growing function (i.e., the eigenstates are stable over the entire existence domain). This peculiarity has been already pointed out in a similar system~\cite{carpentier10}, where the unstable surface modes are suppressed by diminishing the width of the trapping potential. 

One key result of our analysis can be drawn by looking at the dependence of $N$ on the amplitude of the eigenstate at the boundary of nonlinear region $U(x=x_0)$ as shown in Fig.\ref{fig1}(b). It turns out that, at the critical point $N=N_{max}$, the amplitude of the solution at the boundary (black points) approaches a constant value $U_{cr}$ with increasing $x_0$. This indicates that all the solution branches become unstable at the same value $U_{cr}$ for $x_0\gg d$, irrespective of how large the value of $N_{max}$ is. In fact, the maximum number of particles that can be trapped by the linear potential without leading to destabilization of the ultracold atomic cloud increases with growing $x_0$, as it can be appreciated in Fig.\ref{fig1}. We will explain below such features with the aid of a simple analytical model.

\section{Analytical condition for matter-wave emission} 

Interestingly, we observe in Fig. \ref{fig1} that the stable branch of eigenmodes localized inside the potential well lays close to the unstable one corresponding to surface modes for $x_0\gg d$. Then, small perturbations in the system may transform stable solutions into unstable ones, as the coupling between both modes can be accomplished owing to the similarities in both profiles and chemical potentials. Subsequently, the surface modes may eventually decay into stable solitons existing in the nonlinear region. Nevertheless, according to our interpretation of the matter-wave emission, the perturbation should be strong enough to make the amplitude of the solution at the boundary of the nonlinear region greater than critical, namely, $U>U_{cr}$.

\begin{figure}[h]
\epsfig{file=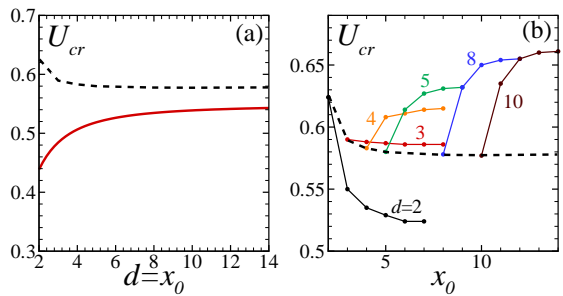,width=8cm}
\caption{ (Color online) Panel (a): Dependence of the critical amplitude $U_{cr}=\left.(U)\right|_{x = x_0}$ on $x_0$ at the point $N=N_{max}$, calculated numerically solving Eq.~(\ref{st_gpe_1}) (black dashed line) and analytical prediction (red solid line) taken from the geometric non-guiding condition (\ref{cond}).  Other parameters are: $V_0=-0.3$, $l=4$. Panel (b): Dependence $U_{cr}(x_0)$ estimated numerically for different widths of the trap $d$. The value of $d$ is indicated for each of the curves. Dashed line corresponds to the condition $x_0=d$.} \label{fig_xcr}
\end{figure}

Let us now support the previous statement by deriving an analytical condition for soliton emission. 
To do so,
we will follow similar arguments to those presented in Ref.~\cite{carpentier10} for obtaining a geometric non-guiding condition. In analogy with integrated optical systems, we consider the linear trapping potential as a one-dimensional linear step-index waveguide of width $2d$, with the core index $V_0$, cladding index $\left. V\right|_{x=-d} = 0$ (left trap border) and substrate index $V_{nl}$ (right border $x_0=d$), $V_{nl}$ being the nonlinear contribution to the effective potential created by particle interactions.
In this framework, the guiding properties of the structure can be fully characterized by both the normalized frequency $f_0 = \sqrt 2 d \sqrt{ V_0 - V_{nl}}$ and the cut-off frequency $f_c(\nu) = \arctan \sqrt{ V_{nl}/(V_0 - V_{nl})}+\nu \pi$, where $\nu$ is a non-negative integer which establishes a cut-off for the existence of guided modes~\cite{tamir}. 
Notice that both $f_0$ and $f_c(\nu)$ are dimensionless, owing to the formal analogy between Eq.~(\ref{st_gpe_1}) and the stationary version of the paraxial Helmholtz equation governing the evolution of light in optical media~\cite{tamir}, which allows to quantify the effective refractive indices $V_0, V_{nl}$ in units of the squared vacuum wavenumber $k_0^2$.

 The particular case $f_c(0)$ corresponds to the threshold of the fundamental mode. For the lowest energy (guided) mode to become a radiating mode, the condition $f_0 < f_c(0)$ has to be satisfied so that particles cannot be trapped inside the guiding potential and will eventually  flow towards the region where inter/intra-species interactions are nonzero. This will happen for values of $V_{nl}$ satisfying  the geometric non-guiding condition $F = f_0 - f_c(0) < 0$.
Hereafter, we will consider $V_{nl}$ to be field-dependent owing to the existence of nonlinear interactions within the
region $x>x_0$. Thus, for the simple scalar case, $V_{nl} = c|U|^2$ at $x=x_0$, showing its explicit dependence on the particle density.

Using the above conditions, the critical amplitude of the condensate wave-function on the right border of the trap at which the threshold of existence occurs (emission condition) can be obtained as
\begin{eqnarray}
U_{cr}=\left.(U)\right|_{x = x_0}>\sqrt{V_{nl}}
\label{cond}
\end{eqnarray}

In Fig.\ref{fig_xcr}(a), we display the explicit comparison between the results obtained from numerical solution of Eq.~(\ref{st_gpe_1}) for the condition $x_0=d$, and those predicted by the simple analytical model (\ref{cond}). As the width of the trap increases, the discrepancy between the two approaches gets reduced [see Fig. \ref{fig_xcr}(a)]. This is because the distortion of the atomic cloud produced by the nonlinear effects is strongly reduced as the mode is mostly confined within the linear trap, and so the accuracy of the analytical model, which assumes a differential overlapping with the nonlinear region, is improved accordingly. Moreover, our model catches another key feature of the system: $U_{cr}$ approaches a constant value as $x_0$ grows, as discussed above. To illustrate quantitatively the accuracy of the model, we have verified that the error between the two curves is below $10\%$ for $d=14$. This deviation can be attributed to the use of a super-Gaussian function with smooth boundaries to 
model $V(x)$ in the numerical solution of Eq.(\ref{st_gpe_1}), whilst for the analytical calculations a step-index waveguide is assumed.

\begin{figure}[h]
\epsfig{file=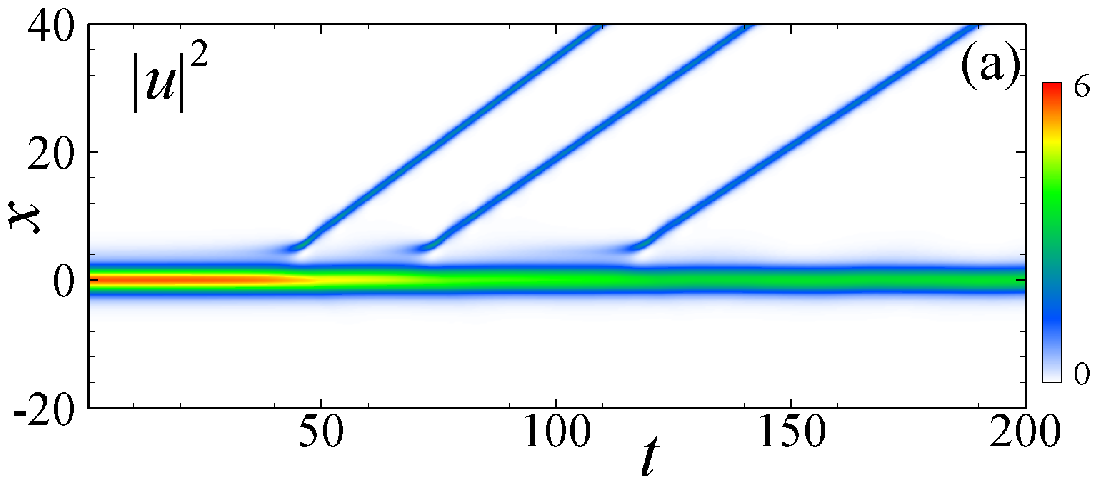,width=8cm}
\epsfig{file=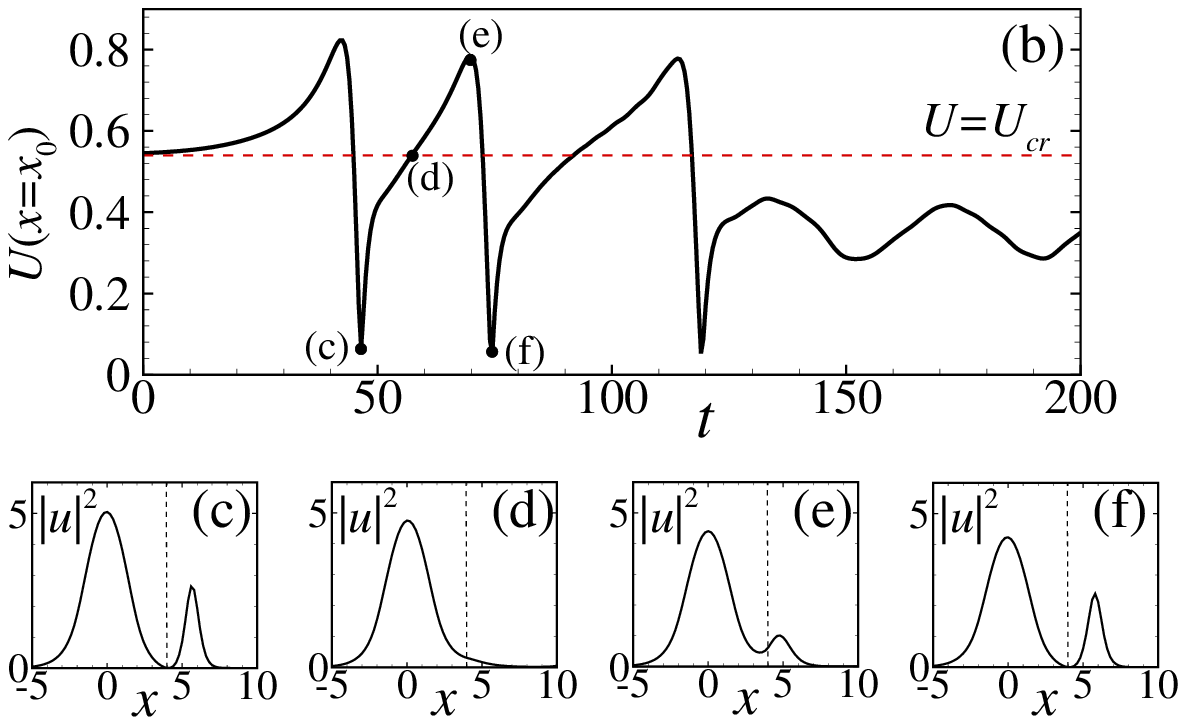,width=8cm}
\caption{(Color online) Panel (a): 
Temporal dynamics of the density $|u(x,t)|^2$ for the eigenstate featuring $N=N_{max}=19.84$ under small perturbations.
The initial condition considered reads $N=(1+\delta)N_{max}$, with $\delta=0.02$. 
Panel (b): Evolution of the amplitude of the solution at the point $x=x_0$. Dashed-red line indicates the outcome of ideal unperturbed dynamics, whereas solid-black line corresponds to the perturbed dynamics shown in (a). 
Panels (c)-(f): profiles of the density of the solution at the points indicated in (b).
The final number of particles localized in the trap is $N_f\approx 11.8$ and the norm of each soliton  is $\approx 2.8$.
Other parameters are: $c=1$, $V_0=-0.3$, $d=2$, $l=4$.} \label{fig1_dyn_max}
\end{figure}

We have checked the validity of the proposed criterion for soliton emission by studying the evolution of the eigenmode for $x_0=4$ at critical point $N=N_{max}$, perturbed by slightly increasing the number of particles, $N=(1+\delta)N_{max}$ with $\delta\ll 1$. The dynamics of the aforementioned atomic cloud is shown in Fig.~\ref{fig1_dyn_max}. As $N>N_{max}$, we observe the launching of solitons from the trapped cloud [see Fig. \ref{fig1_dyn_max}(a)]. Each outcoupled soliton removes a certain amount of particles ($N\approx 2.8$) when  released out of the trap, thus reducing $U(x=x_0)$ and $N$, accordingly. At the same time, it introduces additional perturbations in the system which allows for the critical condition  $U>U_{cr}$ to be dynamically fulfilled, yielding to the outcoupling of a new soliton as long as $N > N_{min}$. In order to get insight into this dynamical phenomenon, we plot the evolution of $U(x=x_0)$ in Fig.~\ref{fig1_dyn_max}(b). We see that the emission of solitons occurs when the amplitude 
$U(x=x_0)$ overcomes the critical amplitude $U_{cr}$. In the panels (c)-(f), we display the snapshots of the density $|u|^2$ at different times during the emission of the second soliton. As it can be appreciated in the graphs, after the outcoupling of the first soliton at $t=46.5$  [panel (c)], the perturbed density of the cloud reaches again the critical value $|U_{cr}|^2$ at the boundary at $t=58$ [panel (d)], thus satisfying the condition for matter-wave emission and triggering the process of formation of the next soliton. At $t=70$, the maximum density at the boundary is achieved [panel (e)] and the outcoupling process continues until the generation of the soliton and contraction of the cloud to its new perturbed eigenstate of the linear trap. Notice that, during this process, the amplitude of the density crosses once again the critical amplitude $U_{cr}$, but in contrast to (d), the crossing at that point happens from the top entering into the undercritical regime at the boundary $U<U_{cr}$. At  $t=74.5$ [panel (f)], the density profile of the atomic cloud resembles that of the eigenstate shown in (c), except for a smaller number of particles.
Remarkably, the launching process stops immediately after emission of the third soliton as the perturbation becomes insufficient to reach $U_{cr}$, i.e., the condition for  soliton emission is never satisfied again [see Fig. \ref{fig1_dyn_max}(b)].

It is relevant to notice that the number of generated solitons grows by increasing $x_0$. The explanation of this relies on the fact that both stable and unstable branches of stationary states become closer in the existence plot $N(\mu)$ [see Fig. \ref{fig1}(a)], and so the coupling between both families of solutions with $N\approx N_{max}$ is favored. Thus, any weak perturbation of the stable eigenmode can trigger the emission of a bright soliton, i.e., in the language of nonlinear guided optics, the mode coupling process can be efficiently phase-matched~\cite{Agrawal, Wright:1990}. That is why in the cases $x_0=2,3$, for example, it is not possible to extract more than one soliton out of the trap due to the large gap between stable and unstable branches.
It should be also mentioned that there are two cases within our parameter space, when soliton emission is not possible considering weak perturbations of the eigenmodes. The first case corresponds to the trap with small width $d<2$, where the unstable branch disappears, i.e.,  the whole existence curve comprises stable solutions. This issue was recently discussed in~\cite{carpentier10}. The second case corresponds to the situation when the initial eigenmode features number of particles $N<N_{min}$, which prevents the excitation of self-sustained bright matter-waves.

Finally, the picture becomes more complex when the potential width is fixed and $x_0$ is allowed to change. We have performed extensive numerical analysis and found that $U_{cr}$ displays a rich behavior depending on $x_0$. Thus, as it can be seen in Fig.\ref{fig_xcr}(b), for $d=2$ ($d>4$) the critical amplitude is a decreasing (increasing) function of $x_0$. This change cannot be properly described by the analytical model introduced at the beginning of this section, since it considers a three-layer system~\cite{tamir} which is no longer the case whenever $x_0>d$.

\section{Soliton trains in binary BECs - Supersolitons}

In the case of two-component condensate the cross-term interactions modify the corresponding effective linear potential at the boundary for the first and second component,
making it difficult to obtain a suitable condition for the emission of solitons. This may eventually complicate the controllable generation of trains of interwoven matter-wave solitons of both atomic species. To circumvent this difficulty, we propose a novel setup to control both the launching condition for each component independently, as well as the period between emission of two consecutive solitons. Thus, let us consider two non-interacting BEC components ($c_{12}=c_{21}=0$), each of them containing a critical number of particles $N_1=N_2=N_{max}$ (the profile shown in Fig.\ref{ini}).  In the absence of any external perturbation, both eigenstates belong to the stable branch and thus the critical condition for soliton emission Eq.~(\ref{cond}) is not overcome. Noteworthy, the particular choice $N_j=N_{max}$ is not strict, and so all the phenomenology to be described below would also hold for $N_j<N_{max}$.

\begin{figure}[h]
\epsfig{file=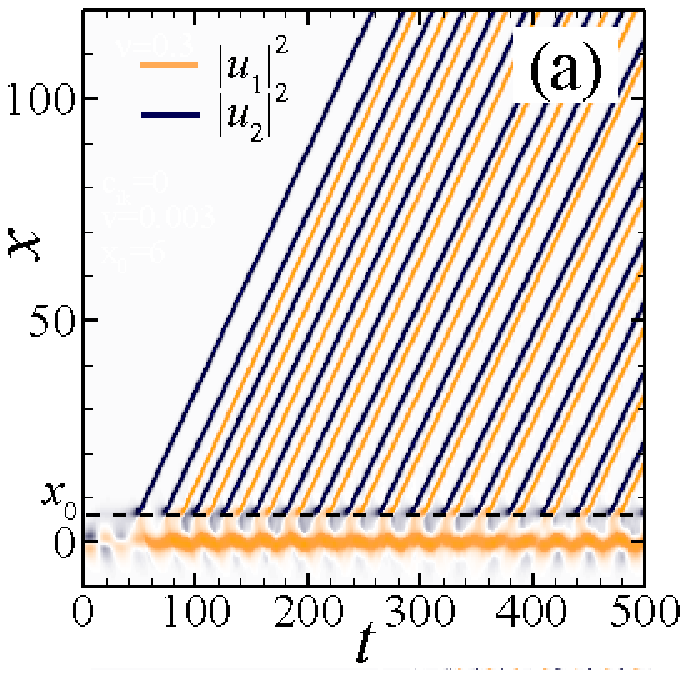,width=4cm}\epsfig{file=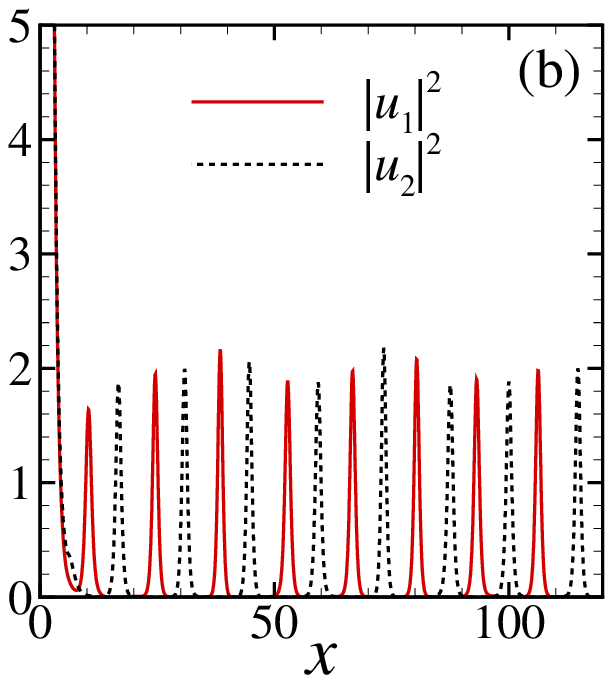,width=4cm}
\caption{(Color online) Panel (a):  density plot of $|u_1|^2-|u_2|^2$ with $c_{jk}=0$. The boundary for intra-species interaction $x_0$ is shown as dashed horizontal line.
Panel (b): density profiles of the components $|u_1|^2$ (red solid line) and $|u_2|^2$ (black dashed line) taken at $t=300$.
Other parameters are: $c_{jj}=1$, $x_{0}=6$, $V_0=-0.3$, $d=2$, $l=4$ and $v=0.003$.} 
\label{train_s12_0}
\end{figure}

In order to alternately launch solitons from the different components, we apply small velocities to the trapped initial atomic clouds. This can be achieved experimentally by phase imprinting technique \cite{phaseimpr}, while in our model it corresponds to just multiplying the condensate wave-functions $u_j$ by a linear phase $e^{i v_j x}$. In particular, we will consider the case of equal (in absolute value) and opposite velocities ($v_1=-v_2$), which results in out-of-phase oscillations of the atomic clouds inside the trap. As shown in Fig.~\ref{fig1_dyn_max}, each time one of the components approaches the nonlinear region at $x=x_0$, it can overcome the critical amplitude $U_{cr}$ and, consequently, a soliton from that component is released. With this method, a regular train of alternating solitons in both components can be easily accomplished (see Fig. \ref{train_s12_0}(a), where we plot the density difference $|u_1|^2-|u_2|^2$ to distinguish the two atomic species). The time delay between 
the emission of solitons from different components approximately corresponds to half of the period of oscillation of the atomic cloud in the trap. Remarkably, all emitted solitons carry almost the same number of particles, featuring slight variations in the amplitudes [see Fig. \ref{train_s12_0}(b)]. 

In addition, for small initial velocities $|v_j|=0.003$, there is a small transient regime for $t<50$ where no soliton emission occurs [see Fig. \ref{train_s12_0}(a)]. This indicates that the critical amplitude $U_{cr}$ is not overcome during the very first oscillations of the atomic clouds, owing to the weak perturbation introduced prior to the dynamics. After some oscillations, and due to the increasing perturbations in the clouds provided by the nonlinear interactions, the soliton emission starts. This effect is analog to the working principle of a passively-driven mode-locked laser \cite{modelocking}.

As we have previously discussed, small initial velocities lead to almost parallel trains of alternating solitons which move with identical velocities as the clouds oscillate in a quasi-periodic manner. For stronger perturbations, however, the dynamics turns out to be more involved.
In fact, the strongly driven system lacks sinusoidal oscillations, resulting in substantial increase in the fluctuation of both soliton amplitudes and velocities, hence distorting the picture of parallel propagation of the outcoupled solitons. We must point out that, also in this case, first solitons from both components are launched without any transient regime, as long as the perturbation is strong enough to overcome $U_{cr}$ along the first oscillation.

Hitherto, we have shown all along this section how to generate trains of interwoven bright solitons in binary BEC by controlling the nonlinear self-interactions. In the following we will explore the effects of nonlinear cross-interactions between solitons belonging to different components. To do so, we will consider $c_{jk}\neq0$ at spatial interval $x_{00}>x_0$ (see Fig.~\ref{ini}). Depending on the sign of the inter-atomic interaction $c_{jk}$, solitons from different components will feel either attractive or repulsive forces. When $c_{jk}=1$, Eq.~(\ref{gpe}) reduces to the integrable Manakov system~\cite{Manakov} well within the nonlinear region, for which solitons are transparent and so emerge unscattered after head-on collisions, except a phase shift. On the other hand, the inclusion of repulsive cross-term interactions with $c_{jk}=-1$ ($j\neq k$) results in the observation of other interesting effects, namely  phase-independent elastic interactions between solitons coming from different components~\cite{novoa08,reflectrans12} 
and eventual generation of SS, i.e. phonon-like excitations living on top of alternating soliton arrays \cite{novoa08}. The experimental scenario proposed in Ref. \cite{novoa08} for the observation of SS was based on the mechanism of modulational instability of two different atomic clouds, coupled to each other via nonlinear cross-interactions switched from repulsive to attractive. A more refined version of this challenging proposal has been recently reported~\cite{feijoo2013}.

\begin{figure}[h]
\epsfig{file=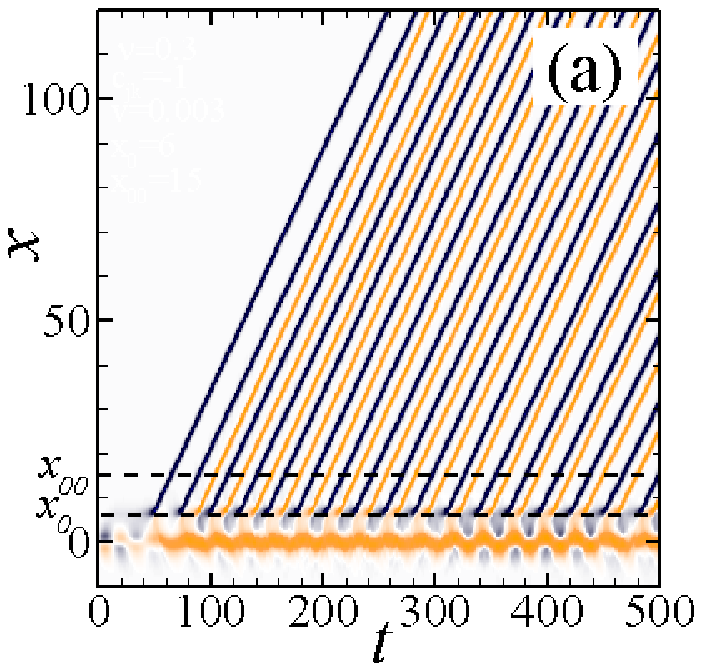,width=4cm}\epsfig{file=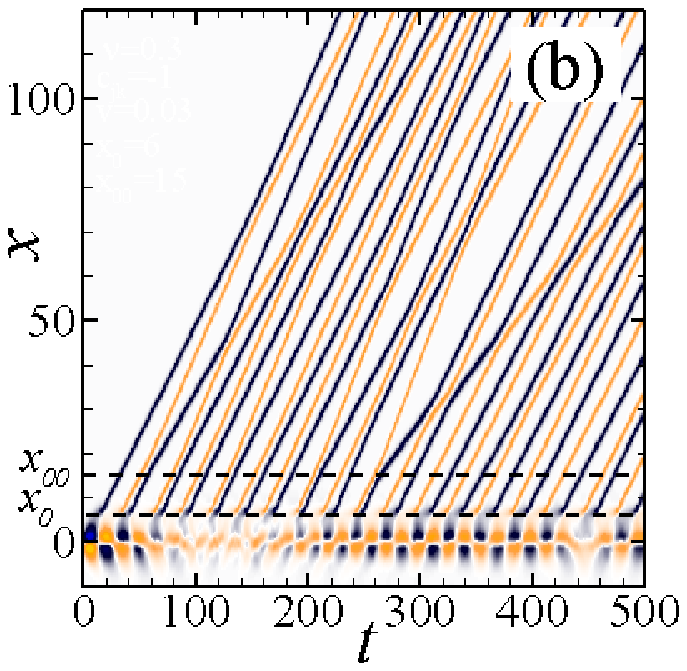,width=4cm}
\caption{ (Color online) Propagation dynamics for (a) weak ($|v|=0.003$) and  (b) strong  ($|v|=0.03$) initial perturbations. Other parameters are the same as in Fig.\ref{train_s12_0} with $c_{jk}=-1$ and $x_{00}=15$.}
\label{train_s12__1_SS}
\end{figure}

Hereafter, we will propose a different scheme to observe the formation of SS. As described above, for small initial velocities ($|v|=0.003$) applied to the atomic cloud, the emitted solitons display almost parallel propagation within the trains. Besides, the overall dynamics of the soliton trains upon propagation does not present any noticeable behavior
as shown in Fig.~\ref{train_s12__1_SS}(a).  In other words, solitons from different components display particle-like (elastic) behavior when they interact in the train, due to the phase-independent interactions~\cite{novoa08}. By increasing the initial perturbation, however, the periodic equidistant launching of the solitons becomes affected and two adjacent
 solitons, ejected from the different components, emerge in the region $x>x_{00}$ with non-zero overlap. This cause them to strongly interact through repulsive inter-species coupling,  
 that turn out to modify the original soliton velocities. Thus, if the change in the  velocity is sufficiently large, solitons can initiate linear momentum transfer to their nearest neighbors through collisions, giving rise to {\it spontaneous emission} of SS. This phenomenon of SS generation (around $t=250$) and its further propagation can be observed in Fig. \ref{train_s12__1_SS}(b). Notice that the velocity of the SS is almost constant in this regime, owing to the onset of elastic interactions between the (equal-mass) quasi-particles.

Nevertheless, even though the spontaneous generation of SS is interesting by itself, one would desire to precisely control their excitation at specific positions and with well-defined velocities, thinking of potential applications. We can achieve this  by changing the velocity of a specific soliton in the train, located at the position where we want to generate the SS. This can be realized by introducing an additional localized repulsive potential in the path of the corresponding soliton; this can be experimentally accomplished by inducing an optical dipole trap with a blue-detuned laser operating outside the main trap~\cite{BECreview,bdlaser1}.

\begin{figure}[h]
\epsfig{file=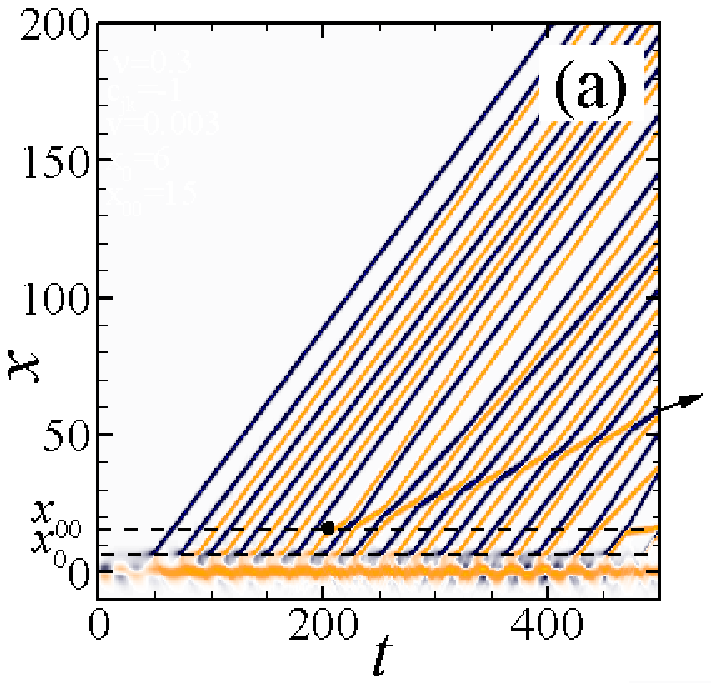,width=4cm}\epsfig{file=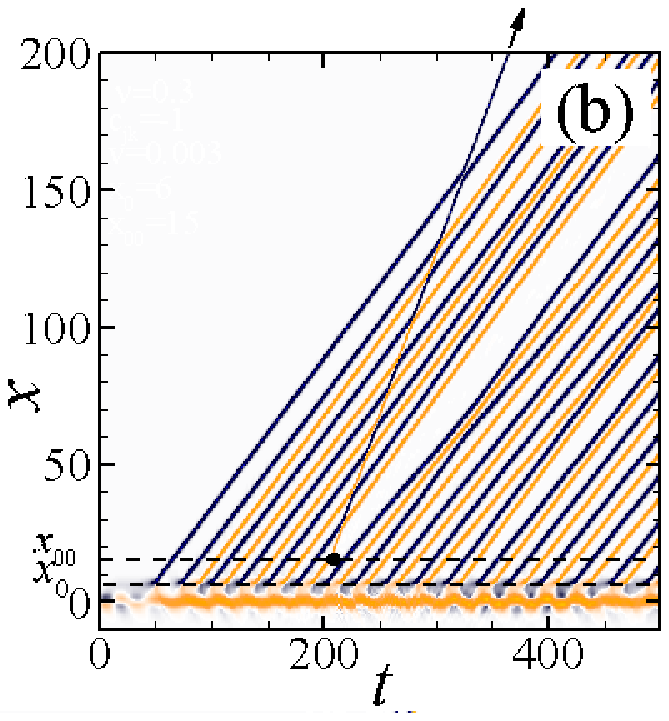,width=4cm}
\caption{(Color online) Generation of slow (a) and fast (b)  SSs in the weakly perturbed configuration shown in Fig.\ref{train_s12__1_SS}(a)  by switching narrow repulsive potentials at  $t_i=196$ with $\tau=6$ in (a)  and $t_i=206$ with $\tau=3$ in (b), respectively.
Parameters of the localized potentials are the following: $\zeta=3$, $x_d=15$ and $\xi=0.25$. The black dots show the time-space position of the external potential $V_g(x)$. The arrows show the directions of the SS's trajectories.} 
\label{slowSS}
\end{figure}

Interaction of the soliton with the repulsive trap modifies the soliton trajectory in a controllable way, either accelerating or decelerating it depending on the relative defect position. We will demonstrate this numerically by generating both slow and fast SS in a weakly perturbed system. We consider a Gaussian localized potential $V_g(x)=\zeta\exp[-(x-x_d)^2/\xi]$  with amplitude $\zeta$ and width $\xi$, which is switched on at the position $x=x_d$ at time $t_i$ with duration $\tau$. We will hereafter treat the dynamically generated interwoven soliton train as being at rest, since all solitons move with almost equal velocities [see Fig. \ref{train_s12__1_SS}(a)]. We do this by introducing the following Galilean transformation $x'\rightarrow x-v_t t$, where the co-moving frame $x'$ travels at the velocity of the soliton train, which in our calculation corresponds to $v_t\approx 0.54$. The latter procedure allows us to rigorously define the velocity of the SS with respect to the soliton train.

Then, when the repulsive potential is placed in the path of a certain soliton, this gets decelerated and interacts elastically with other solitons in the train. As a result, a slow SS with velocity $v_s\approx 0.14$ is excited, as it can be appreciated in Fig.\ref{slowSS}(a). Similarly, switching on the potential just behind one of the solitons in the train results in the acceleration of that solitary wave, and subsequently gives rise to the generation of a fast SS with velocity $v_f \approx 1.14$ [see Fig.\ref{slowSS}(b)]. 

\begin{figure}[h]
\epsfig{file=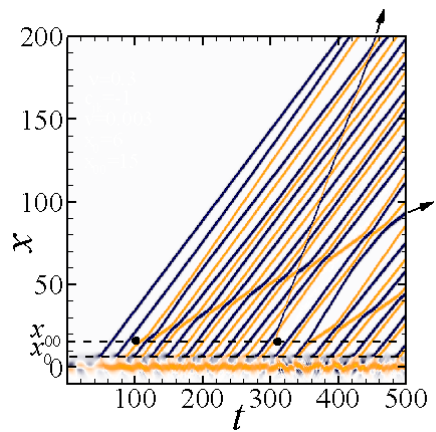,width=6cm}
\caption{ (Color online) Head-on collision of slow and fast SS generated by switching on narrow repulsive potentials at  $t_i=99$ with duration $\tau=2.3$ and $t_i=306$ with duration $\tau=2$. Other parameters are the same as in  Fig.\ref{slowSS}.} 
\label{twoSS}
\end{figure}

Finally, in order to prove that these phonon-like localized excitations display indeed solitonic nature, we have studied their dynamics in a head-on collision interaction. In Fig. \ref{twoSS} we observe that by first generating a slow SS and then forcing its collision with a fast SS, generated at a later time, the two localized excitations pass through each other without changing their velocities, as expected for solitons in effectively integrable systems.

\section{Conclusions}

We have studied the dynamics of binary Bose-Einstein condensates under spatial variation of the intra- and inter-component interactions between ultracold atoms. We have put forward a method for the generation of trains of alternating bright solitons, using both controlled emission of nonlinear matter-waves from each of the components in the uncoupled regime and out-of-phase oscillations of the ground states in the trap. Bright solitons from different components are then sequentially outcoupled from the linear potential. They interact with each other through the repulsive cross-coupling (switched on after formation of solitons) resulting in total phase-independent (classical-like) soliton interactions. By including additional localized external potentials in the model, we can excite fast and slow supersoliton that are shown to behave like solitary waves in integrable systems. 
The above mechanism of controllable generation of supersoliton is very general, allowing us to consider further extension of the study including manipulation of supersoliton and their management, as well as generalization to higher-dimensional systems.
The present results on the generation and control of the dynamics of interwoven soliton trains could also have implications in matter-wave interferometry \cite{interferometer} and realization of effective supersolid-like structures with ultracold atoms~\cite{supersolid}.

\acknowledgments

D. N. acknowledges support from MINECO through FCCI. ACI-PROMOCIONA (ACI2009-1008) and Consolider program SAUUL (CSD2007-00013). J.C.P. acknowledges support from FCT Grant No. SFRH/BPD/77524/2011.


\begin{thebibliography}{99}

\bibitem{anderson95} M. H. Anderson, J. R. Ensher, M. R. Matthews, C. E. Wieman and E. A. Cornell, Science {\bf 269}, 198 (1995); K. B. Davis, M. O. Mewes, M. R. Andrews, N. J. van Druten, D. S. Durfee, D. M. Kurn and W. Ketterle, \prl {\bf 75}, 3969 (1995).

\bibitem{interferometer}  M. A. Kasevich, C.R. Acad. Sci. IV {\bf 2},  497 (2001).

\bibitem{michinel12} H. Michinel, A. Paredes, M. M. Valado, and D. Feijoo, \pra {\bf 86}, 013620 (2012).

\bibitem{carpentier10} A. V. Carpentier, H. Michinel, D. N. Olivieri and D. Novoa, J. Phys. B: At. Mol. Opt. Phys. {\bf 43}, 105302 (2010).

\bibitem{mewes97} M. O. Mewes, M. R. Andrews, D. M. Kurn, D. S. Durfee, C. G. Townsend and W. Ketterle, \prl {\bf 78}, 582 (1997).

\bibitem{carr04} L. D. Carr and J. Brand, \pra {\bf 70}, 033607 (2004).

\bibitem{rodas05} M. I. Rodas-Verde, H. Michinel and V. M. P\'erez-Garc\'{i}a, \prl {\bf 95}, 153903 (2005); 
A. V. Carpentier, H. Michinel, M. I. Rodas-Verde and V. M. P\'erez-Garc\'{i}a, \pra {\bf 74}, 013619 (2006).

\bibitem{guerin06} W. Guerin, J. F. Riou, J. P. Gaebler, V. Josse, P. Bouyer and A. Aspect, \prl {\bf 97}, 200402 (2006).

\bibitem{couvert08} A. Couvert, M. Jeppesen, T. Kawalec, G. Reinaudi, R. Mathevet and D. Guery-Odelin, Europhys. Lett. {\bf 83}, 50001 (2008).

 \bibitem{gattobigio09} G. L. Gattobigio, A. Couvert, M. Jeppesen, R. Mathevet and D. Guery-Odelin, \pra {\bf 80}, 041605 (2009).

\bibitem{solitonBEC} L. Khaykovich, F. Schreck, G. Ferrari, T. Bourdel, J. Cubizolles, L. D. Carr, Y. Castin and C. Salomon, Science {\bf 296}, 1290 (2002).

\bibitem{perez98} V. M. P\'erez-Garc\'{i}a, H.  Michinel and H. Herrero, \pra {\bf 57}, 3837 (1998).

\bibitem{train02} K. E. Strecker, G. B. Partridge, A. G. Truscott and R. G. Hulet, Nature {\bf 417}, 150 (2002).

\bibitem{thalhammer08} G. Thalhammer, G. Barontini, L. De Sarlo, J. Catani, F. Minardi and M. Inguscio, \prl {\bf 100}, 210402 (2008).

\bibitem{Best09} Th. Best, S. Will, U. Schneider, L. Hackerm\"uller, D. van Oosten, I. Bloch and D. S. L\"uhmann, \prl {\bf 102}, 030408 (2009).

\bibitem{vector_solitons} F. V. Pepe, P. Facchi, G. Florio and S. Pascazio \pra {\bf 86}, 023629 (2012);
A. I. Yakimenko, K. O. Shchebetovska, S. I. Vilchinskii and M. Weyrauch \pra {\bf 85}, 053640 (2012);
L. Wen, W. M. Liu, Y. Cai, J. M. Zhang and J. Hu, \pra {\bf 85}, 043602 (2012); 
Z. M. He, D. L. Wang, J. W. Ding and X. H. Yan, Eur. Phys. J. D. {\bf 66}, 139 (2012);
J. Stockhofe, P. G. Kevrekidis, D. J. Frantzeskakis and P. Schmelcher, J. Phys. B: At. Mol. Opt. Phys. {\bf 44}, 191003 (2011);
C. Yin, N. G. Berloff, V. M. P\'erez-Garc\'{i}a, D. Novoa, A. V. Carpentier and H. Michinel, \pra {\bf 83}, 051605(R) (2011). 

\bibitem{novoa08} D. Novoa, B. A. Malomed, H. Michinel and V. M. P\'erez-Garc\'{i}a, \prl {\bf 101}, 144101 (2008).

\bibitem{feijoo2013} D. Feijoo, A. Paredes and H. Michinel, \pra  {\bf 87}, 063619 (2013).

\bibitem{dipoletrap} R. Grimm, M. Weidem\"uller and Y. N. Ovchinnikov, Adv. At. Mol. Opt. Phys. {\bf 42}, 95 (2000).

\bibitem{BECreview} A. J. Leggett, \rmp {\bf 73}, 307 (2001).

\bibitem{Inouye98} S. Inouye, M. R. Andrews, J. Stenger, H. J. Miesner, D. M. Stamper-Kurn and W. Ketterle, Nature {\bf 392}, 151 (1998).

\bibitem{Fatemi00} F. K. Fatemi, K. M. Jones and P. D. Lett, \prl {\bf 85}, 4462 (2000). 

\bibitem{submicron} R. Yamazaki, S. Taie, S. Sugawa, Y. Takahashi, \prl {\bf 105}, 050405 (2010). 

\bibitem{Akhmediev82} N. N. Akhmediev, Zh. Eksp. Teor. Fiz. {\bf 83}, 545 (1982).

\bibitem{VK} N. G. Vakhitov and A. A. Kolokolov, Rad. Quant. Elect. {\bf 16}, 783 (1973).

\bibitem{Agrawal} G. P. Agrawal, \emph{Nonlinear Fiber Optics - Third Edition}, Ed. Academic Press, London (2001).

\bibitem{Wright:1990} E. M. Wright, D. R. Heatley and G. I. Stegeman, Phys. Rep. {\bf 194}, 309 (1990).

\bibitem{tamir} T. Tamir,  \emph{Integrated Optics}, Springer, Berlin (1979).

\bibitem{phaseimpr} J. Denschlag \emph{et al.}, Science {\bf 287}, 97 (2000).

\bibitem{modelocking} O. Svelto, \emph{Principles of Lasers - Fifth Edition}, Ed. Springer, New York (2010).

\bibitem{Manakov} S.A. Manakov, Sov. Phys. JETP {\bf 38}, 248 (1974).

\bibitem{reflectrans12} H. Zhang-Ming, W. Dong-Long, D. Jian-Wen and Y. Xiao-Hong, Commun. Theor. Phys. {\bf 58}, 381 (2012). 

\bibitem{bdlaser1} X. He, S. Yu, P. Xu, J. Wang and M. Zhan, Opt. Express {\bf 20}, 3711 (2012).

\bibitem{supersolid} L. Mathey, I. Danshita and C. W. Clark, \pra {\bf 79}, 011602(R) (2009).


\end{thebibliography}
\end{document}